\documentclass[sigconf]{acmart}

\usepackage{amsmath}
\usepackage{bm}
\usepackage{algorithm}
\usepackage{algpseudocode}
\usepackage{amsmath}
\usepackage{multirow}
\usepackage{indentfirst}
\usepackage{subfigure}
\usepackage{epsfig}
\usepackage{epstopdf}
\usepackage{graphicx}
\usepackage{url}
\usepackage{xspace}
\usepackage{booktabs}
\usepackage{subeqnarray}
\usepackage{subcaption}
\usepackage{color}

\newcommand{\paratitle}[1]{\vspace{1.5ex}\noindent\textbf{#1}}
\newcommand{\ie}{\emph{i.e.,}\xspace}
\newcommand{\aka}{\emph{a.k.a.,}\xspace}
\newcommand{\eg}{\emph{e.g.,}\xspace}

\newcommand{\ignore}[1]{}

\AtBeginDocument{%
  }

\begin{document}

\title{Holistically Guided Monte Carlo Tree Search for Intricate Information Seeking}


\author{Ruiyang Ren$^*$}\thanks{$^*$Equal Contributions.}
\email{reyon_ren@outlook.com}
\affiliation{%
  \institution{Gaoling School of Artificial Intelligence, Renmin University of China}
  \city{Beijing}
  \country{China}}

\author{Yuhao Wang$^*$}
\email{yh.wang500@outlook.com}
\affiliation{%
  \institution{Gaoling School of Artificial Intelligence, Renmin University of China}
  \city{Beijing}
  \country{China}}

\author{Junyi Li$^*$}
\email{junyi_cs@nus.edu.sg}
\affiliation{%
  \institution{National University of Singapore}
  \country{Singapore}}

\author{Jinhao Jiang}
\email{jiangjinhao@ruc.edu.cn}
\affiliation{%
  \institution{Gaoling School of Artificial Intelligence, Renmin University of China}
  \city{Beijing}
  \country{China}}

\author{Wayne Xin Zhao$^\dag$}\thanks{$^\dag$Corresponding Authors.}
\email{batmanfly@gmail.com}
\affiliation{%
  \institution{Gaoling School of Artificial Intelligence, Renmin University of China}
  \city{Beijing}
  \country{China}}

\author{Wenjie Wang$^\dag$}
\email{wenjiewang96@gmail.com}
\affiliation{%
  \institution{University of Science and Technology of China}
  \city{Hefei}
  \country{China}}

\author{Tat-Seng Chua}
\email{dcscts@nus.edu.sg}
\affiliation{%
  \institution{National University of Singapore}
  \country{Singapore}}

\renewcommand{\shortauthors}{Ren et al.}

\begin{abstract}

In the era of vast digital information, the sheer volume and heterogeneity of available information present significant challenges for intricate information seeking. Users frequently face multi-step web search tasks that involve navigating vast and varied data sources. This complexity demands every step remains comprehensive, accurate, and relevant. However, traditional search methods often struggle to balance the need for localized precision with the broader context required for holistic understanding, leaving critical facets of intricate queries underexplored.
In this paper, we introduce an LLM-based search assistant that adopts a new information seeking paradigm with holistically guided Monte Carlo tree search (HG-MCTS). We reformulate the task as a progressive information collection process with a knowledge memory and unite an adaptive checklist with multi-perspective reward modeling in MCTS. 
The adaptive checklist provides explicit sub-goals to guide the MCTS process toward comprehensive coverage of complex user queries. Simultaneously, our multi-perspective reward modeling offers both exploration and retrieval rewards, along with progress feedback that tracks completed and remaining sub-goals, refining the checklist as the tree search progresses. By striking a balance between localized tree expansion and global guidance, HG-MCTS reduces redundancy in search paths and ensures that all crucial aspects of an intricate query are properly addressed.
Extensive experiments on real-world intricate information seeking tasks demonstrate that HG-MCTS acquires thorough knowledge collections and delivers more accurate final responses compared with existing baselines.

\end{abstract}

\maketitle

\section{Introduction}

In real-world web search systems, addressing an information seeking task often requires retrieving and organizing information from diverse online sources. This task that we term \emph{intricate information seeking} presents a persistent and significant challenge in the field of information retrieval~\cite{strohman2005optimization, talmor2018web}. Unlike conventional single-step search, where a user query typically seeks isolated information, intricate information seeking involves integrating multiple pieces of information across various sources to formulate a comprehensive and accurate final response.
The complexity of this process is further amplified by the necessity to maintain consistency across retrieval steps, especially when user queries encompass multifaceted tasks or require extensive background knowledge. For example, responding to a complex query such as ``\emph{the economic, environmental, and social impacts of the adoption of renewable energy in developing countries}'' entails sourcing multiple relevant documents and synthesizing them into a comprehensive answer, including analyzing economic benefits and costs, assessing environmental sustainability, evaluating social implications.

Typically, driven by the ongoing evolution of large language models~(LLMs)~\cite{zhao2023survey}, a variety of existing methods aim to facilitate multi-step or complex retrieval either by heuristically decomposing the query or by iteratively refining the query through incremental optimization of intermediate outputs~\cite{yao2023react, asaiself}. For instance, some studies adopt planning strategies by decomposing a user query into sub-queries based on surface-level cues with the LLM's internal knowledge~\cite{xu2024search, reddy2024infogent}, while others employ tailored mechanisms~(\eg chain-of-thought reasoning~\cite{wei2022chain} and continuous feedback loops~\cite{shinn2024reflexion}) to progressively align intermediate reasoning steps with the final information seeking goal. Although these methods have shown promise in improving multi-step retrieval quality, they are susceptible to cascading errors, where inaccuracies or omissions in earlier sub-queries can propagate through subsequent steps. 

Inspired by the effectiveness of Monte Carlo Tree Search (MCTS) applied in complex reasoning tasks such as mathematic and code problems~\cite{alphago, YeLKAG21}, we consider incorporating MCTS into intricate information seeking scenarios to help find the optimal retrieval solution. However, two primary challenges emerge when applying MCTS to the task. First, sub-queries generated for expanding the search tree are unbounded at each step: each multifaceted intermediate task can branch into numerous investigative directions, causing the tree search space to grow exponentially. Second, the inherent characteristics of MCTS lead to a focus on local exploration~\cite{browne2012survey}, which may lead to omissions or solecism in the acquired information~\cite{_wiechowski_2022}. Specifically, (1) MCTS node selection relies on local statistics (\eg number of visits and reward accumulation), which are aggregated from limited exploration and lack a holistic understanding of the global information seeking objectives; and (2) MCTS explores the search tree by expanding nodes around the currently selected branch, while its rollout strategies are typically random or heuristic-based, making global optimality difficult to guarantee. Consequently, MCTS risks overlooking pertinent sub-tasks or prematurely converging on suboptimal search paths.

\begin{figure}
    \centering
    \includegraphics[width=0.95\linewidth]{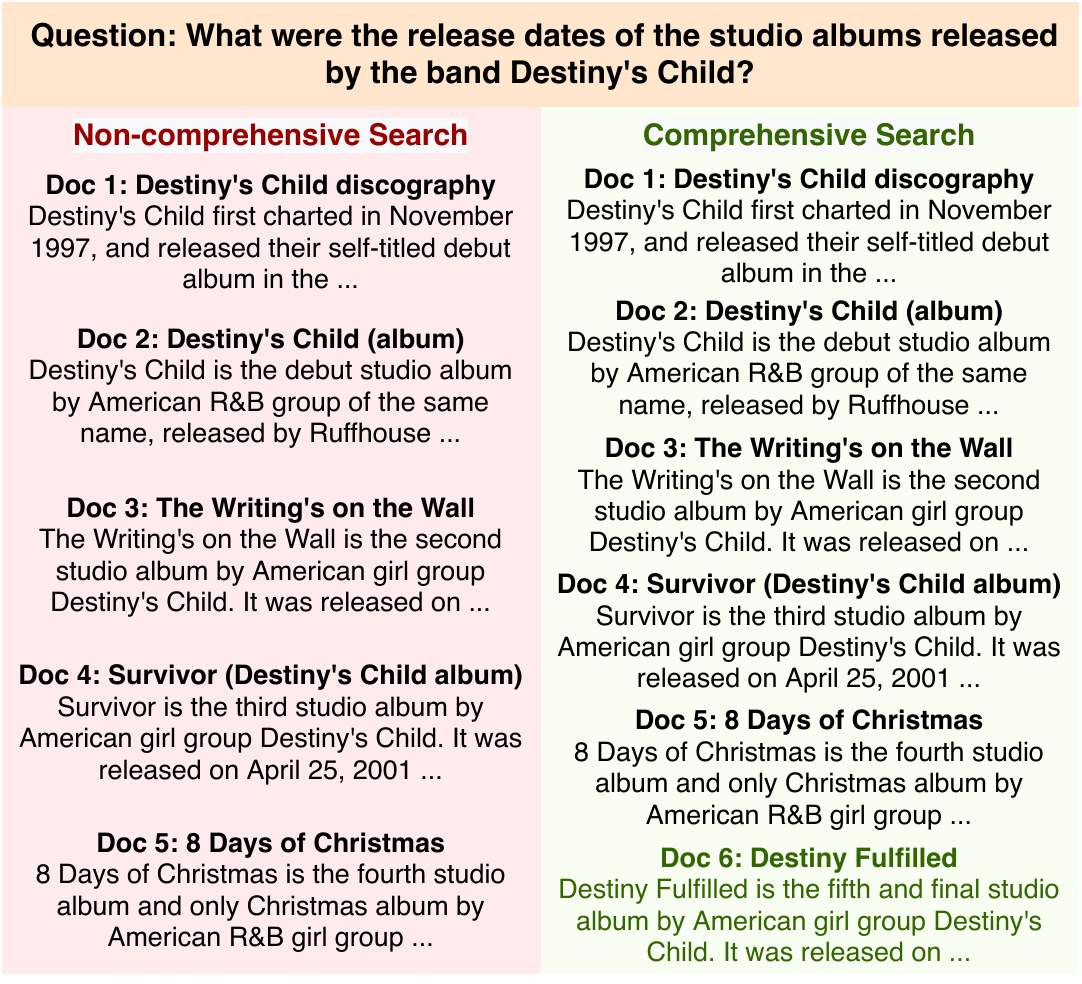}
    \caption{Illustration of the pitfalls in handling intricate queries. Typical reasoning methods with web search often collect non-comprehensive documents (left), while HG-MCTS can effectively capture all necessary documents (right).}
    \label{fig:intro}
\end{figure}

To address these challenges, we propose a novel framework that incorporates MCTS into intricate information seeking, while simultaneously mitigating its inherent limitations through global guidance and multi-perspective feedback. 
Concretely, we reformulate the task as a progressive information collection process with a knowledge memory. Based on this, we propose \emph{holistically guided MCTS~(HG-MCTS)} that introduces an \emph{adaptive checklist} as a global guidance with a set of designated sub-goals.
This adaptive checklist counters the exponential growth of sub-queries by focusing the MCTS algorithm on only those branches aligned with key facets of the information need, thereby alleviating aimless expansions that could arise and enforcing a more targeted traversal of the search space, which can also be updated during the MCTS process. In parallel, we incorporate \emph{multi-perspective reward modeling} that provides both quantitative and qualitative reward signals with the checklist, allowing MCTS to incorporate a more holistic perspective on exploration. This reward modeling furnishes not just numerical indicators of exploration and retrieval quality but also textual feedback outlining which sub-goals have been addressed and which remain unsolved after node exploration. As a result, MCTS moves beyond its conventional reliance on local statistics, thereby minimizing the risk of prematurely converging on suboptimal paths and broadening its understanding of overarching information seeking objectives. Figure~\ref{fig:intro} illustrates a comparison between our method and the typical information seeking method from the retrieval comprehension perspective. 
Through this synergy, our approach preserves the inherent capability of MCTS for dynamic exploration while strengthening its capacity to incorporate newly acquired knowledge snippets. Our method methodically balances thoroughness and focus, ensuring comprehensive coverage of all relevant information while avoiding redundant or tangential searches.

Our main contributions are summarized as follows:
\begin{itemize}
    \item We introduce a new information seeking paradigm \emph{HG-MCTS} based on a progressive information collection process with knowledge memory, which integrates an adaptive checklist for holistic sub-goal guidance in MCTS progress, thereby enabling more targeted exploration in multi-step retrieval.
    \item We propose a \emph{multi-perspective reward modeling} strategy that provides both quantitative metrics and qualitative feedback in HG-MCTS, which fosters a richer, step-wise evaluation for the value of new expanded nodes.
    \item We demonstrate how these innovations can be seamlessly integrated to improve both the efficiency and the thoroughness of large-scale web retrieval. Beyond immediate applications in question answering and knowledge-intensive search, our findings offer deeper insights into the design of more interpretable, flexible, and resilient retrieval systems.
\end{itemize}

\begin{figure*}[t]
    \centering
    \includegraphics[width=0.999\linewidth]{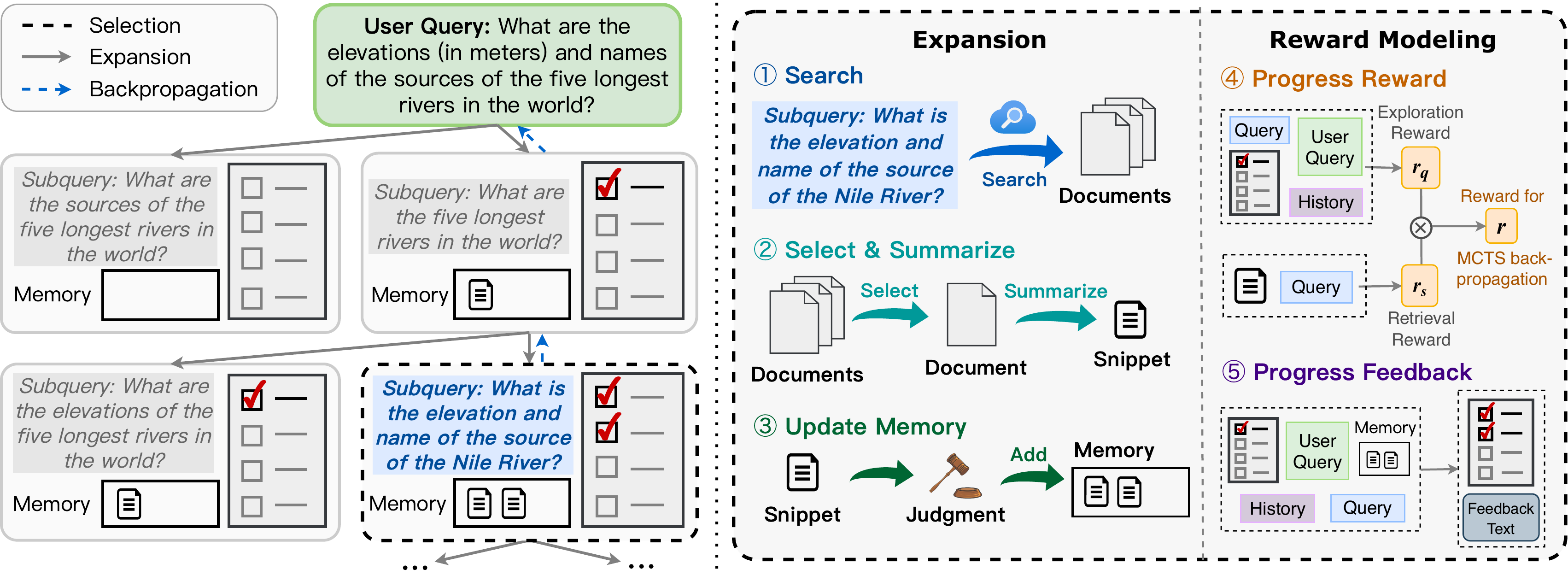}
    \caption{The overall framework of the proposed HG-MCTS method. The left panel outlines the iterative Monte Carlo tree search procedure with a global checklist and memory mechanism, including different actions in MCTS. The right panel provides a detailed explanation of node expansion and the corresponding reward modeling process with quantitative progress reward and progress feedback.}
    \label{fig:framework}
\end{figure*}

\section{Preliminaries}

We begin by formally defining the task of intricate information seeking, then present the Monte Carlo Tree Search~(MCTS) algorithm based on large language models (LLMs), and provide the formulation of our MCTS-driven information seeking task.

\paratitle{Intricate Information Seeking.} The intricate information seeking task involves the systematic exploration and retrieval of information from various sources to address complex and multifaceted queries. Formally, given a query $\bm{x}$, the goal of search assistant $\mathcal{M}_\theta$ is to retrieve a set of documents $\mathcal{D}=\{d_1,...,d_K\}$ and generate an answer $\bm{y} = \langle y_1,...,y_m \rangle$ based on the retrieved information, where $y_t$ denotes a token from a pre-defined vocabulary $\mathcal{V}$. Compared to traditional web search tasks where queries are often straightforward and focused on specific pieces of information (\eg ``\emph{current weather in London}''), the queries in intricate information seeking involve multiple dimensions or subtopics that require aggregating information from various aspects (\eg ``\emph{the medal standings of the past five Summer Olympic}''). Besides, our task emphasizes gathering a wide range of documents to provide exhaustive coverage of topics, requiring retrieving data from diverse types of sources (\eg government reports and factual articles). Based on the accessed information, intricate information seeking aims to incorporate the main content from multiple sources, integrating facts, insights, and diverse viewpoints to ensure that the final answer is sufficiently accurate and comprehensive to address the query.

\paratitle{Monte Carlo Tree Search.} Monte Carlo Tree Search is a search technique for possible solutions in the field of artificial intelligence. It combines the principles of tree exploration and random simulation to estimate the potential outcomes of actions. MCTS is especially effective in decision-making tasks with large action spaces, such as AlphaGO~\cite{alphago} and Atari~\cite{YeLKAG21}. 
In complex reasoning tasks, previous work (\eg Chain-of-Thoughts~\cite{wei2022chain} and ReAct~\cite{yao2023react}) mainly focused on chain-based reasoning paradigm, where any intermediate reasoning errors may lead to incorrect final answers. Furthermore, although some studies incorporated MCTS into retrieval~\cite{lee2024zero,jiang2024rag}, they simply apply RAG in each expansion step and search for the most effective intermediate answers aiming to derive the final answer, which may become trapped in locally optimal search paths or overlook critical pieces of required information.
In contrast, we reformulate the task as an information collection process that mainly focus on collecting accurate and comprehensive information with MCTS and utilizing the extracted information stored in the memory to generate the final answer. Specifically, we introduce an adaptive checklist to provide global guidance for the exploration and reward modeling in MCTS. By imposing atomic sub-goals on the tree search process, our approach systematically advances toward the overall goal. It effectively minimizes ineffective searches and ensures the comprehensive collection of all necessary information. We also incorporate progress feedback to update the checklist and provide the next direction. Consequently, the final answers become more precise and are accompanied by accurate retrieval references, thereby enhancing the explainability of the information seeking process.

\paratitle{MCTS-based Information Seeking.} Inspired by previous studies on complex reasoning~\cite{kang2024mindstar, wang2024q*, jiang2024technical}, we propose to integrate MCTS to deal with the intricate information seeking task. In our study, MCTS aims to build a search tree $\mathcal{T}$ based on a policy model $\pi_{\theta}$, which is usually the target LLM $\mathcal{M}_\theta$. To provide holistic guiding signals, we incorporate a global checklist $p$ listing sub-goals for addressing the input complex query.
In the search tree $\mathcal{T}$, a node $s_t = [q_t, p, \mathcal{D}]$ denotes the state at the $t$-th tree level, where $q_t$ is an intermediate sub-query for retrieval, $p$ is the dynamically updated checklist indicating solved and unsolved sub-goals, and $\mathcal{D}$ denotes the current knowledge memory containing a set of extracted knowledge from retrieved documents so far. Based on the policy model, we sample candidate actions from its probability distribution $\mathcal{M}_\theta(a_t|s_t)$, transiting from the current state $s_t$ to the next state $s_{t+1}$. The MCTS process begins from a root node $s_0=[\bm{x}]$, as the initial input query, and iterates four key steps (\ie selection, expansion, evaluation, and backpropagation) to finally access a comprehensive set of knowledge memory $\mathcal{D} = \{k_1,..., k_T$\}, where $k_t$ denotes knowledge snippet extracted from retrieved documents at the $t$-th step and $T$ is the number of MCTS iterations. Based on the collected information $\mathcal{D}$, the policy model $\mathcal{M}_\theta$ generates the final answer $\bm{y}$ to the input query.

\section{Methodology}

In this section, we introduce the proposed approach \underline{H}olistically \underline{G}uided \underline{M}onte \underline{C}arlo \underline{T}ree \underline{S}earch, named \emph{HG-MCTS}, for the intricate information seeking task by leveraging exploration and exploitation capabilities of the MCTS algorithm.

\subsection{Overall Framework}

To tackle the challenges of the intricate information seeking task, we propose an integrated framework named HG-MCTS. Based on the reformulated information collecting based reasoning process, HG-MCTS seamlessly combines the capabilities of \emph{adaptive checklist} and \emph{multi-perspective reward modeling} to guide the MCTS process for targeted exploration on pre-planned retrieval sub-goals. Figure~\ref{fig:framework} conceptually illustrates the proposed framework.

The HG-MCTS framework begins by taking an \emph{intricate user query} as input. Leveraging the policy model, it generates an \emph{Adaptive Checklist}, consisting of anticipated sub-goals for resolving the query. This list of sub-goals constitutes an explicit guiding signal, significantly narrowing the expansive search space and enabling the MCTS to focus on promising retrieval sub-goal paths. Once the checklist is established, MCTS repeatedly performs the four canonical steps of \emph{selection}, \emph{expansion}, \emph{evaluation}, and \emph{backpropagation}. During \textbf{selection}, the node with the highest UCT value is chosen; in \textbf{expansion}, the policy model proposes a specialized query by incorporating both the \emph{adaptive checklist} and a running \emph{progress feedback} $u$ from the reward modeling. 
The search engine is subsequently queried to retrieve new documents, after which the policy model selects one document to extract key information in the form of snippets. This extracted knowledge is then assessed for inclusion in the existing \emph{knowledge memory} $\mathcal{D}$, ensuring that the information acquired at each step is systematically retained for subsequent iterations.

Another core of our approach lies in its \emph{multi-perspective reward modeling} (Section~\ref{sec-evaluation}) in \textbf{evaluation} phase, which provides comprehensive assessment and detailed progress feedback during the MCTS process. Specifically, the reward model produces two quantitative measures: the \emph{exploration reward}, which assesses whether the newly proposed query aligns well with unaddressed sub-goals, and the \emph{retrieval reward}, which measures how effectively the contained knowledge in the retrieved documents address the information need of the newly proposed subquery. These reward signals are aggregated and propagated in the MCTS tree in \textbf{backpropagation} phase, updating the value function of each visited node. In tandem with these numerical rewards, the \emph{progress feedback} furnishes textual hints about completed and remaining sub-goals in the adaptive checklist. This complementary reward modeling mechanism enables the policy model to dynamically revise the search strategies to explore more efficient reasoning paths.

Under our holistically guided approach, the MCTS algorithm persistently refines its node value with tailored rewards and explores new possible directions, ensuring coverage of all sub-goals. With the adaptive checklist, the MCTS process naturally terminates when the progress feedback indicates that all sub-goals on the checklist have been addressed. Finally, the knowledge snippets stored in the knowledge memory comprehensively cover every sub-goal of the user query and are consolidated to produce the final response. In this manner, {HG-MCTS} achieves a synergy between explicit goal specification and targeted exploration, offering an efficient and effective solution for intricate information seeking.

\begin{figure}
    \centering
    \includegraphics[width=0.8\linewidth]{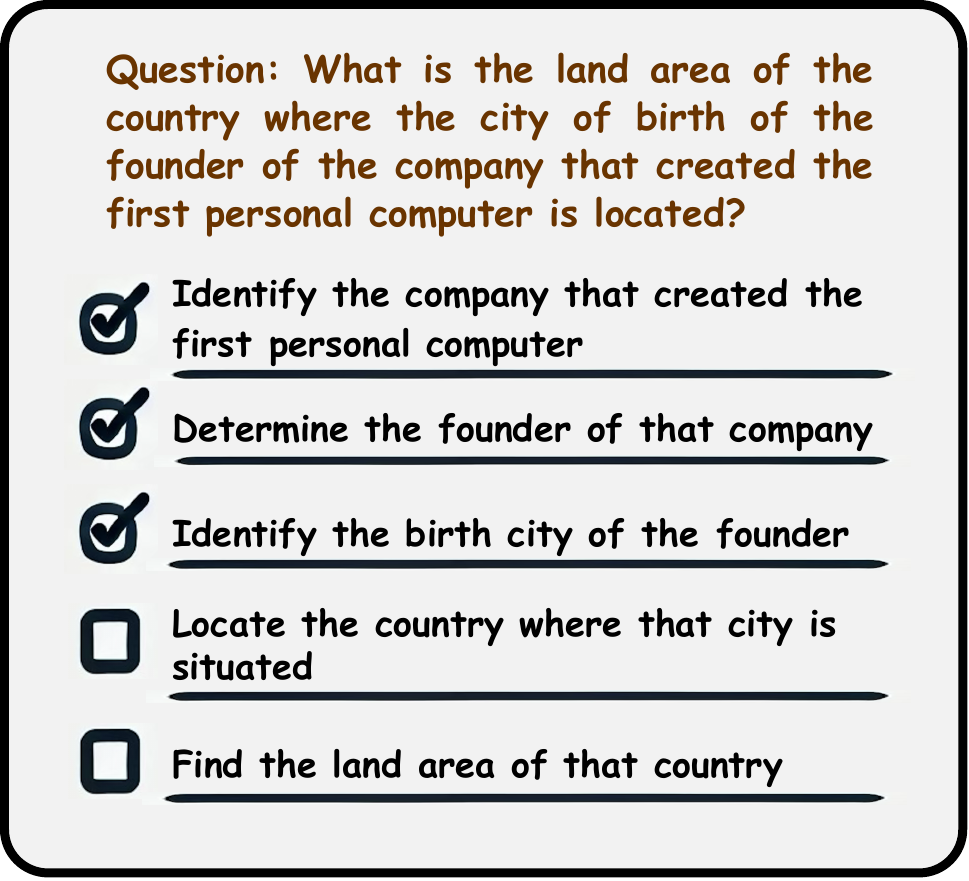}
    \caption{An illustration of the checklist corresponding to an intricate query encompassing multiple sub-goals. 
    }
    \label{fig:checklist}
\end{figure}

\subsection{Holistically Guided Monte Carlo Tree Search}

To achieve holistic exploration for intricate information seeking, we first employ the LLM to generate an adaptive checklist as a holistic signal for MCTS, which then iteratively performs selection, expansion, evaluation, and backpropagation based on the LLM.

\subsubsection{Checklist Generation.} Previous MCTS studies usually rely on the MCTS algorithm to traverse all possible actions for deriving the optimal solution~\cite{alphago, YeLKAG21}, which will lead to massive search space and substantial computational costs. Therefore, we propose to incorporate \emph{Adaptive Checklist}, an explicit guiding signal that outlines all sub-goals required to solve the multifaceted queries in intricate information seeking. This checklist serves to direct the MCTS algorithm toward unsolved sub-goals, effectively avoiding meaningless action searches and thereby improving the overall search efficiency. Specifically, based on the input query $\bm{x}$, the policy model $\mathcal{M}_\theta$ first generates a checklist $p = \langle w_1,...,w_n \rangle $ in the form of lists relying on its understanding and internal knowledge as follows:
\begin{equation}
    w_{i+1} = \mathcal{M}_\theta(\bm{x}, \{w_1,...,w_i\}),
\end{equation}
where $w_{i+1}$ denotes the $(i+1)$-th token in the checklist text. Figure~\ref{fig:checklist} presents an illustrative example for the checklist in our approach. After generating a checklist as the global signal, the policy model iterates the four steps in MCTS, which are detailed in the following. 

\subsubsection{Selection}
In this process, the MCTS algorithm traverses the current tree starting from the root node using a specific strategy, which adopts a scoring function to optimally select nodes with the highest estimated value. The Upper Confidence Bound applied to Trees (UCT) formula~\cite{uct} is usually employed as the strategy in the selection process to traverse the tree. It balances the exploration-exploitation trade-off during tree traversal by selecting a node with the highest UCT score as below:
\begin{equation}
    UCT(s_t) = V(s_t) + w\sqrt{\frac{\ln{N(s_{t-1})}}{N(s_t)}},
\end{equation}
where $V(s_t)$ is the value of node $s_t$ which will be estimated at the evaluation process, $N(s_t)$ denotes the visit count of node $s_t$, $s_{t-1}$ is the parent node of $s_t$, and $w$ is a hyper-parameter controlling the exploration and exploitation between these two parts.

\subsubsection{Expansion}
After selecting a node $s_t$ with the highest UCT score, it will be expanded by generating multiple child nodes $\{s_{t+1}\}$. In our method, the policy model expands a new node in two steps, \ie firstly propose an intermediate subquery $q_{t+1}$ that aims to address certain sub-goals in the checklist and secondly retrieve some relevant documents from external sources (\eg web) via the proposed subquery. Compared to previous work~\cite{quiet-star,zhang2024llama} that expands new nodes only relying on historical states, we consider incorporating the adaptive checklist $p$ to generate more accurate and targeted subqueries $q$ as follows:
\begin{equation}
    q_{t+1} = \mathcal{M}_\theta(\mathcal{H}, p, \mathcal{D}),
\end{equation}
where the history context $\mathcal{H} = \{ \bm{x}, u, q_1,..., q_t \}$ contains the complex input query $\bm{x}$, last progress feedback $u$~(detailed in Section~\ref{sec:feedback}) and previous generated subqueries $q_1,..., q_t$, and $\mathcal{D}$ denotes the knowledge memory from the last state. In this formula, the checklist $p$ describes a list of completed and remaining unsolved sub-goals and steps, serving as a holistic search signal for addressing the input query.
This mechanism enables the MCTS algorithm to explore the space of unresolved sub-goals more efficiently, thereby avoiding redundant searches. Based on the proposed subquery $q_{t+1}$, the policy model then leverages search engine (\eg Google Search) to retrieve a set of documents $\mathcal{C}_{t+1}$ from the web. Since the primary objective of intricate information seeking is to gather and aggregate all necessary information as comprehensively as possible, we maintain a knowledge memory $\mathcal{D}$ throughout the entire MCTS process to store the collected knowledge snippets. After retrieving new documents $\mathcal{C}_{t+1}$ via $q_{t+1}$, the policy model will filters and incorporates only the useful information into the memory. To conserve memory space, we adopt an abstractive approach by selectively choosing the most relevant document and summarizing it to obtain the necessary content $k_{t+1}$, then the knowledge is subsequently evaluated and incorporated into the memory $\mathcal{D}$:
\begin{equation}
    \mathcal{D} \xleftarrow[]{\text{add}} k_{t+1} \xleftarrow[]{\text{summarize}} \mathcal{M}_\theta(\mathcal{C}_{t+1}),
\end{equation}
where $k$ represents the extracted knowledge content from the retrieved documents $\mathcal{C}_{t+1}$ by the policy model.

\subsubsection{Evaluation}
The evaluation process aims to compute the expected reward $r$ for the newly expanded nodes $s_{t+1}$ using the multi-perspective reward modeling (detailed in Section~\ref{sec-evaluation}). On the one hand, the reward model first computes an \emph{exploration reward} $r_q$ evaluating whether the generated subquery $q_{t+1}$ corresponds to specific unfinished sub-goals in the checklist and assessing its degree of global consistency based on the historical context information from the root node to the current node, \ie $\mathcal{H} = \{ \bm{x}, u, q_1, ..., q_t \}$. On the other hand, the reward model provides a \emph{retrieval reward} $r_k$ to evaluate the relevance of the retrieved knowledge snippet $k_{t+1}$. The final reward for the expanded node sums up the two aspects of rewards: 
\begin{equation}
    r = r_q \cdot r_k. 
    \label{eq:reward}
\end{equation}
We find that numerical rewards alone are insufficient to effectively guide the policy model in exploring the vast search space. Therefore, the reward model is further designed to provide a \textit{progress feedback} $u$ based on the adaptive checklist $p$, historical information $\mathcal{H}$, and the current state $s_{t+1}$. The progress feedback text will be returned to the policy model to update the checklist for indicating which goals in the checklist have been completed and which remain unresolved as follows:
\begin{equation}\label{eq-update-checklist}
    p =\text{Update}(p, u; \mathcal{M}_\theta).
\end{equation}

\subsubsection{Backpropagation} 
After evaluating the value of the newly expanded node, the remaining tree must be updated. So, the backpropagation process is performed, where the reward $r$ is propagated back from the new node $s_{t+1}$ to the root node $s_0$. During the process, the number of visit counts for each node is incremented by one and the node value will be accordingly updated as follows:
\begin{align}
    N_{\text{new}}(s_j) &= N_{\text{old}}(s_j) + 1,~~0 \leq j \leq t, \\
    V_{\text{new}}(s_j) &= \frac{V_{\text{old}}(s_j)N_{\text{old}}(s_j) + r}{N_{\text{new}}(s_j)},
\end{align}
where $N_{\text{old}}(s_j)$ and $V_{\text{old}}(s_j)$ represent the last visit count and value of node $s_j$ before backpropagation, respectively, and $r$ is the reward obtained from the evaluation process.

\subsubsection{Answer Generation}

By performing the above four steps iteratively, the MCTS process terminates until reaching the maximum tree level or the progress feedback indicates that the sub-goals in the checklist have been all addressed, so that it obtains a comprehensive knowledge memory $\mathcal{D}=\{k_1,...,k_T\}$. Based on the collected information, the policy model generates the final answer as follows:
\begin{equation}
    y_j = \mathcal{M}_\theta(\mathcal{D}, \{y_1,...,y_{j-1}\}),~~1 \leq j \leq m,
\end{equation}
where $y_j$ is the $j$-th token in the answer.

\subsection{Multi-Perspective Reward Modeling}
\label{sec-evaluation}

Traditional MCTS approaches usually perform expensive rollout operations to the current state until the terminal state evaluates the expected reward of the current state~\cite{alphago,YeLKAG21}. Inspired by previous work on process-supervised reward modeling~\cite{setlur2024rewarding,LightmanKBEBLLS24}, we propose directly leveraging an external reward model to provide step-by-step rewards from multiple perspectives, \ie exploration reward, retrieval reward, and progress feedback in Equation~(\ref{eq:reward}).

\subsubsection{Exploration Reward}
To evaluate the overall quality of the proposed intermediate subquery $q_{t+1}$ generated by the policy model, we propose to employ the reward model $\mathcal{R}_\theta$ to provide the exploration reward $r_q$.
This reward evaluates the subquery's consistency with any unfinished sub-goal in the checklist $p$, intricate user query $\bm{x}$ and assesses its degree of global consistency based on the historical context information from the root node to the current node:
\begin{equation}
    r_{q} = \mathcal{R}_\theta(q_{t+1}, p, \mathcal{H}).
\end{equation}
The exploration reward $r_q$ is an integer, either 0 or 1. An output of 0 indicates that, after comprehensively considering the given information, the exploration direction of the current intermediate subquery is deemed suboptimal. Conversely, an output of 1 signifies that the subquery is appropriate for acquiring the next piece of information.
The exploration reward assesses the directional information acquisition of the policy model at each step during the MCTS process, thereby providing effective guidance for identifying more optimal reasoning paths.

\subsubsection{Retrieval Reward}
To evaluate the comprehensiveness and relevance of atomic knowledge contents in the knowledge memory $\mathcal{D}$, we propose to employ the reward model $\mathcal{R}_\theta$ to compute the retrieval reward $r_k$:
\begin{equation}
    r_{k} = \mathcal{R}_\theta(q_{t+1}, k_{t+1}).
\end{equation}
We constrain $r_{k}$ to a three-tier integer scoring system, corresponding to three levels of semantic relevance between the atomic knowledge content $k_{t+1}$ the current subquery $q_{t+1}$:
\begin{equation}
r_k=\left\{
\begin{aligned}
0, \quad &\text{if}~k_{t+1}~\text{cannot satisfy the information need of}~q_{t+1}& \\
1, \quad &\text{if}~k_{t+1}~\text{partially satisfy the information need of}~q_{t+1}& \\
2, \quad &\text{if}~k_{t+1}~\text{fully satisfy the information need of}~q_{t+1}&
\end{aligned}
\right. \nonumber
\end{equation}
The retrieval reward assesses the policy model’s actions in searching for and extracting relevant information, facilitating the discovery of reasoning paths with enhanced information acquisition quality.

\subsubsection{Progress Feedback}
\label{sec:feedback}
Numerical rewards alone may not effectively guide the policy model through the extensive search space. Therefore, the reward model is further designed to deliver a \textit{progress feedback} $u$, based on the subquery $q_{t+1}$, adaptive checklist $p$, historical information $\mathcal{H}$, and knowledge memory $\mathcal{D}$. The progress feedback $u$ obeying the text form, which can be generated as:
\begin{equation}
    u = \mathcal{R}_\theta(q_{t+1}, p, \mathcal{H}, \mathcal{D}).
\end{equation}
Progress feedback delineates which sub-goals in the checklist have been achieved and which remain unresolved, thereby offering clearer direction for the policy model's exploration. The progress feedback will be used to update the adaptive checklist in Equation~(\ref{eq-update-checklist}).
Based on the statistical monitoring of each sub-goal’s completion status, progress feedback can issues an explicit termination signal upon the fulfillment of all sub-goals, thereby concluding the current search process.

Furthermore, certain queries resist decomposition into atomic information-gathering sub-goals, leading to instances where the checklist cannot explicitly define sub-goals prior to the reasoning. For example, the question ``What were the release dates of the studio albums released by the band Destiny’s Child?'' cannot ascertain the required number of search steps without first determining the total number of Destiny’s Child studio albums. To address this limitation, the progress feedback enables the checklist to be {refined dynamically} during the reasoning process based on newly acquired information, thereby accommodating queries that require additional information to establish an adaptive checklist.

\begin{table*}[t]
\centering
\small
\scalebox{1}{
\begin{tabular}{lccccccccccccccc}
\toprule
\multicolumn{1}{c}{\multirow{2.5}{*}{\textbf{Method}}} & \multicolumn{3}{c}{{HotpotQA}} & \multicolumn{3}{c}{{2WikiMultihopQA}} & \multicolumn{3}{c}{{MuSiQue}} & \multicolumn{3}{c}{{StrategyQA}} & \multicolumn{3}{c}{\textbf{Average}} \\
\cmidrule(r){2-4} \cmidrule(r){5-7} \cmidrule(r){8-10} \cmidrule(r){11-13} \cmidrule(r){14-16}
\multicolumn{1}{c}{} & \textbf{EM} & \textbf{CEM} & \textbf{F1} & \textbf{EM} & \textbf{CEM} & \textbf{F1} & \textbf{EM} & \textbf{CEM} & \textbf{F1} & \textbf{EM} & \textbf{CEM} & \textbf{F1} & \textbf{EM} & \textbf{CEM} & \textbf{F1} \\

\midrule
\multicolumn{16}{c}{\textit{GPT-4o-mini (Closed-source)}} \\
\midrule
Closed-Book & 28.82 & 39.80 & 34.71 & 24.71 & 30.40 & 24.71 & 7.65 & 15.91 & 9.41 & 73.53 & 73.53 & 73.53 & 33.68 & 39.91 & 35.59 \\
\multicolumn{1}{l}{Chain-of-Tought} & 26.47 & 38.79 & 32.35 & 24.12 & 30.84 & 26.47 & 13.53 & 20.92 & 18.24 & 51.76 & 52.09 & 51.76 & 28.97 & 35.66 & 32.21 \\
\multicolumn{1}{l}{Standard RAG} & 41.18 & 54.36 & 52.94 & 25.88 & 32.09 & 27.65 & 11.76 & 19.91 & 16.47 & 58.24 & 59.53 & 58.24 & 34.27 & 41.47 & 38.83 \\
\cmidrule{2-16}
\multicolumn{1}{l}{ReAct} & 35.88 & 51.08 & 42.35 & 29.41 & 35.46 & 30.00 & 10.00 & 18.05 & 12.35 & 36.47 & 40.46 & 36.47 & 27.94 & 36.26 & 30.29 \\
\multicolumn{1}{l}{Query2doc} & 44.71 & 57.21 & 54.71 & 29.41 & 34.46 & 29.41 & 19.41 & 28.05 & 24.71 & 64.71 & 65.67 & 64.71 & 39.56 & 46.35 & 43.39 \\
\multicolumn{1}{l}{Self-RAG} & 38.82 & 50.32 & 47.65 & 26.47 & 31.87 & 27.65 & 13.53 & 21.29 & 16.47 & 68.82 & 69.10 & 68.82 & 36.91 & 43.15 & 40.15 \\
\cmidrule{2-16}
Ours & 41.76 & 45.88 & 58.69 & 52.94 & 65.75 & 53.53 & 23.67 & 33.21 & 26.04 & 77.67 & 77.67 & 77.67 & \ \ \textbf{49.01}$^{\dagger}$ & \ \ \textbf{55.63}$^{\dagger}$ & \ \ \textbf{53.98}$^{\dagger}$ \\

\midrule
\multicolumn{16}{c}{\textit{Gemini-1.5-flash (Closed-source)}} \\
\midrule
Closed-Book & 19.41 & 31.52 & 24.71 & 24.12 & 30.07 & 24.71 & 3.53 & 10.98 & 6.47 & 32.35 & 32.83 & 32.35 & 19.85 & 26.35 & 22.06 \\
\multicolumn{1}{l}{Chain-of-Tought} & 28.82 & 36.43 & 34.71 & 18.24 & 21.35 & 18.82 & 8.24 & 13.73 & 10.59 & 68.82 & 69.64 & 68.82 & 31.03 & 35.29 & 33.24 \\
\multicolumn{1}{l}{Standard RAG} & 37.65 & 48.95 & 47.06 & 16.47 & 20.54 & 17.65 & 7.06 & 11.54 & 10.00 & 52.94 & 53.89 & 52.94 & 28.53 & 33.73 & 31.91 \\
\cmidrule(lr){2-16}
\multicolumn{1}{l}{ReAct} & 34.71 & 45.44 & 42.35 & 18.24 & 22.33 & 18.82 & 5.88 & 10.38 & 7.06 & 34.91 & 36.03 & 34.91 & 23.44 & 28.55 & 25.79 \\
\multicolumn{1}{l}{Query2doc} & 38.16 & 49.15 & 47.81 & 17.06 & 20.32 & 17.65 & 11.18 & 17.65 & 14.71 & 58.24 & 58.53 & 58.24 & 31.16 & 36.41 & 34.60\\
\multicolumn{1}{l}{Self-RAG} & 39.83 & 49.47 & 47.88 & 15.29 & 19.06 & 17.65 & 7.65 & 12.09 & 10.00 & 54.12 & 54.53 & 54.12 & 29.22 & 33.79 & 32.41 \\
\cmidrule(lr){2-16}
Ours & 47.65 & 57.65 & 61.98 & 62.94 & 62.94 & 73.94 & 17.75 & 21.30 & 28.31 & 76.19 & 76.19 & 76.19 & \ \ \textbf{51.13}$^{\dagger}$ & \ \ \textbf{54.52}$^{\dagger}$ & \ \ \textbf{60.11}$^{\dagger}$ \\
\midrule
\midrule
\multicolumn{16}{c}{\textit{DeepSeek-V3-chat (Open-source)}} \\
\midrule
Closed-Book & 35.88 & 48.58 & 44.12 & 35.88 & 41.85 & 37.06 & 11.18 & 19.54 & 14.71 & 64.71 & 65.10 & 64.71 & 36.91 & 43.77 & 40.15 \\
\multicolumn{1}{l}{Chain-of-Tought} & 38.82 & 50.59 & 47.06 & 47.06 & 56.14 & 48.24 & 21.18 & 29.99 & 24.12 & 41.76 & 47.46 & 41.76 & 37.21 & 46.12 & 40.30 \\
\multicolumn{1}{l}{Standard RAG} & 40.00 & 55.01 & 50.00 & 30.00 & 34.26 & 31.18 & 15.88 & 24.47 & 18.82 & 64.12 & 65.59 & 64.12 & 37.50 & 44.83 & 41.03 \\
\cmidrule(lr){2-16}
\multicolumn{1}{l}{ReAct} & 40.00 & 54.97 & 48.24 & 32.35 & 36.17 & 32.35 & 16.67 & 34.07 & 20.83 & 23.53 & 28.79 & 23.53 & 28.14 & 38.50 & 31.24 \\
\multicolumn{1}{l}{Query2doc} & 47.19 & 63.11 & 58.05 & 32.35 & 37.14 & 32.35 & 21.18 & 29.89 & 24.71 & 57.65 & 59.40 & 57.65 & 39.59 & 47.39 & 43.19 \\
\multicolumn{1}{l}{Self-RAG} & 43.49 & 56.31 & 51.75 & 27.65 & 32.27 & 28.24 & 16.86 & 25.30 & 19.77 & 52.87 & 53.66 & 52.87 & 35.22 & 41.89 & 38.16 \\
\cmidrule{2-16}
\textbf{Ours} & 45.88 & 52.94 & 62.99 & 60.00 & 63.53 & 73.15 & 21.30 & 32.60 & 25.44 & 76.47 & 76.47 & 76.47 & \ \ \textbf{50.91}$^{\dagger}$ & \ \ \textbf{56.39}$^{\dagger}$ & \ \ \textbf{59.51}$^{\dagger}$ \\
\bottomrule
\end{tabular}}
\caption{The evaluation results for four representative multi-hop QA datasets are presented, we also report the average results of the four datasets. The symbol ``$^{\dagger}$'' denotes that the performance improvement is statistically significant with p < 0.05 compared against all the baselines.}
\label{tab:main-result}
\end{table*}

\section{Experiments and Analysis}
In this section, we begin by detailing the experimental setup, followed by a comprehensive presentation of the primary results. Subsequently, we conduct an ablation study and provide an in-depth analysis to elucidate our findings further.

\subsection{Experimental Settings}

\subsubsection{Datasets}
To rigorously assess the efficacy of our proposed methodology, we conducted evaluations using 5 benchmark datasets. All datasets are meticulously designed to challenge models with complex, multi-hop questions that require synthesizing information across multiple documents.

\begin{itemize}
    \item \textbf{FanOutQA}~\cite{zhu2024fanoutqa} is a high-quality dataset comprising complex information-seeking questions and human-written decompositions, which necessitate aggregating information about multiple entities from several sources to formulate a comprehensive answer. 

    \item \textbf{HotpotQA}~\cite{yang2018hotpotqa} is a widely used dataset for multi-hop question answering, designed to evaluate reason abilities across multiple documents to answer complex questions. The dataset is collected via crowdsourcing based on Wikipedia articles, and annotators are asked to propose questions that require reasoning using the multiple presented Wikipedia articles.

    \item \textbf{2WikiMultihopQA}~\cite{ho2020constructing} is a large-scale multi-hop QA dataset that requires reading multiple paragraphs to answer a given question. The dataset includes four types of questions: comparison, inference, compositional, and bridge-comparison. Each question is accompanied by relevant Wikipedia paragraphs as evidence.
    
    \item \textbf{MuSiQue}~\cite{trivedi2022musique} is created by composing questions from multiple existing single-hop datasets. The dataset is more challenging than previous multi-hop reasoning datasets, with a threefold increase in the human-machine gap and significantly lower disconnected reasoning scores, indicating reduced susceptibility to shortcut reasoning.
    
    \item \textbf{StrategyQA}~\cite{geva2021did} focuses on open-domain questions that require implicit reasoning steps. The dataset consists of 2,780 examples, each comprising a strategy question, its decomposition, and evidence paragraphs. Each question is accompanied by decomposed reasoning steps and relevant Wikipedia paragraphs as evidence.

\end{itemize}

\subsubsection{Evaluation Metrics}
We adopted the established evaluation metrics for the adopted datasets to ensure consistency and comparability. For the evaluation of \textit{FanOutQA}, we employed string accuracy, which measures the proportion of exact matches, and ROUGE metrics~\cite{lin2004rouge}, which assess the quality of summarization by comparing multiple features between the generated and reference texts. Specifically, we report ROUGE-1 (R-1), ROUGE-2 (R-2), and ROUGE-L (R-L) scores to comprehensively assess performance.
For \textit{HotpotQA, 2WikiMultihopQA, MuSiQue}, and \textit{StrategyQA}, we adopt Exact Match (EM), F1 score, and Cover Exact Match (CEM) as evaluation metrics. EM measures strict correctness by checking if the predicted answer matches the ground truth. F1 evaluates the overlap between prediction and ground truth, balancing precision and recall. CEM extends EM to multi-hop reasoning, requiring both correct answers and coverage of intermediate reasoning steps.
Similar to the setup in the existing work~\cite{jiang2024rag}, due to the large data scale, we randomly sampled 130 queries from each of the four datasets for evaluation.

\subsubsection{Baselines}
In the evaluation of our proposed method, we compare it against abundantly established baselines to ensure a comprehensive understanding of its performance. These baselines represent a spectrum of approaches commonly employed in the field, ranging from vanilla reasoning strategies to advanced reasoning methods.

\paratitle{Vinilla reasoning.}\quad
The \textit{Closed-Book} method directly prompts the LLM to provide an answer to a question. In contrast, \textit{Chain-of-Thought (CoT)}~\cite{wei2022chain} reasoning involves adding intermediate reasoning steps to facilitate the response. \textit{Standard RAG} first retrieves passages from the Wikipedia corpus using DPR~\cite{dpr2020} and then directly prompts the LLM to refer to these passages in its response. 

\paratitle{Advanced reasoning.}\quad
\textit{ReAct}~\cite{yao2023react} progressively addresses subqueries, ultimately consolidating the intermediate results to form a complete answer.
\textit{Query2doc}~\cite{wang2023query2doc} generates an initial answer using the model and subsequently retrieves relevant information to generate the final answer.
\textit{Self-RAG}~\cite{asaiself} involves first retrieving information and then assessing its relevance before deciding whether to incorporate it into the final answer.
\textit{MindSearch}~\cite{chen2024mindsearch} employs a planner-searcher architecture for searching relevant information.
\textit{Infogent}~\cite{reddy2024infogent} introduces a multi-agent architecture to aggregate multi-source information.

To ensure a more comprehensive evaluation of different methods, and to mitigate the influence of any specific model, we employ multiple LLMs as backbone models of different methods, including the closed-source LLM \texttt{GPT-4o-mini}, \texttt{Gemini-1.5-flash-002} and the open-source LLM \texttt{deepseek-v3-chat}.
We evaluated all baseline methods in a \textit{zero-shot} setting, employing them solely for inference without additional training.
Note that certain methods are challenging to replicate across all datasets due to the requirement of dataset-specific refinement. As a result, we are unable to report the results for all baselines on every dataset.

\subsubsection{Implement Details}

For web search, we employed Google Search as the primary search engine, selecting the top-3 web search results as document candidates and adhering to existing methodologies for web crawling and denoising~\cite{reddy2024infogent}. During the MCTS process, we ensured consistency between the policy model and the reward model. The maximum number of simulations was capped at 40, and the search depth was limited to 6 layers. In the upper confidence bound for trees algorithm, the exploration-exploitation balance parameter \( w \) was set to 0.2. Additionally, we generated three sub-queries per iteration (\( m_q = 3 \)).
For all generation tasks, responses were sampled using a temperature of 0.9 and top-\( p \) sampling with \( p = 1.0 \).  All prompts used are shown in the provided anonymous codes.

\subsection{Main Results}
The results of different methods evaluated on five complex reasoning datasets are shown in Table~\ref{tab:main-result} and Table~\ref{tab:fanoutqa}. It can be observed that:

(1) Our proposed method demonstrates significant improvements over all baseline approaches across four multi-hop QA datasets and the FanoutQA dataset that emphasizes the information-gathering task. This performance advantage is consistently observed across multiple popular backbone LLMs, highlighting the general applicability and effectiveness of our HG-MCTS framework. HG-MCTS employs an adaptive checklist to guide the expansion and reward modeling of Monte Carlo Tree Search (MCTS), effectively curtailing the exploration of unproductive pathways while maintaining robust search capabilities. In addition, its emphasis on targeted information gathering minimizes irrelevant content, reducing extraneous noise that could compromise the quality of the generated answers.

(2) We further observe that the MCTS-based approach for complex reasoning outperforms both single-step reasoning and chain-based reasoning methods, indicating that tree search can substantially expand the search space of solution paths, enable the discovery of optimal reasoning paths, and ultimately enhance performance. In addition, incorporating advanced reasoning strategies into RAG proves superior to vanilla reasoning standard RAG or closed-book approaches overall, highlighting the importance of meeting users’ complex information needs through tailored retrieval processes. Notably, on the FanoutQA dataset, our method based on GPT-4o-mini surpasses the baseline built upon the larger and more powerful LLM GPT-4o, offering further evidence for the intrinsic advantages of our approach.

(3) Finally, our method exhibits strong robustness across different backbone LLMs, whereas other methods often experience significant performance fluctuations depending on the underlying model. For example, when using Gemini as the backbone, the performance of baseline methods drops markedly compared with results obtained using the other two LLMs, while our method does not exhibit performance degradation. Additionally, we observe that the open-source model deepseek-v3-chat demonstrates capabilities comparable to its proprietary counterparts on various challenging multi-hop QA tasks, thereby laying a promising foundation for real-world deployment of related methodologies.

\begin{table}[t]
    \centering
    \small
\scalebox{1.015}{
    \setlength{\tabcolsep}{3.3pt} 
    \begin{tabular}{llcccc}
        \toprule
        \textbf{Method} & \textbf{LLM} & \textbf{Acc.} & \textbf{R-1} & \textbf{R-2} & \textbf{R-L} \\
        \midrule
        Closed-Book & LLaMA3 & 46.60 & 46.30 & 26.40 & 38.70 \\
        Closed-Book & GPT-4o & 44.10 & 47.40 & 27.30 & 41.70 \\
        Standard RAG & LLaMA3 & 46.80 & 28.20 & 14.30 & 24.30 \\
        Standard RAG & GPT-4o & 58.00 & 49.40 & 31.00 & 44.30 \\
        MindSearch & GPT-4o-mini & 47.30 & 49.30 & 28.40 & 44.20 \\
        Infogent & GPT-4o-mini & 51.10 & 53.30 & 33.00 & 48.50 \\
        \midrule
        \textbf{Ours} & GPT-4o-mini & \ \ \textbf{58.38}$^{\dagger}$ & \ \ \textbf{55.02}$^{\dagger}$ & \ \ \textbf{35.45}$^{\dagger}$ & \ \ \textbf{49.40}$^{\dagger}$ \\
        \bottomrule
    \end{tabular}
    }
    \caption{The evaluation results on FanoutQA. The symbol ``$^{\dagger}$'' denotes that the performance improvement is statistically significant with p < 0.05 compared against all the baselines. LLaMA3 is the abbreviation of LLaMA3-70B-Instruct.}
    \label{tab:fanoutqa}
\end{table}

\subsection{Ablation Studies}
\label{sec:ablation}

In this section, we conduct an ablation study to validate the effectiveness of key strategies in HG-MCTS comprehensively on FanoutQA. Here, we consider five variants based on HG-MCTS for comparison: (a) \underline{\textit{w/o Exploration Reward}} removes the exploration reward during reward modeling; (b) \underline{\textit{w/o Retrieval Reward}} removes the retrieval reward from the total reward during reward modeling; (c) \underline{\textit{w/o Progress Feedback}} eliminates generating the progress feedback in reward modeling; (d) \underline{\textit{w/o Checklist}} removes checklist for global guidance in the MCTS process; (e) \underline{\textit{w/o HG-MCTS}} removes the entire HG-MCTS strategy, reducing the approach to a linear reasoning strategy.

Table~\ref{tab:ablation} presents the results for the variants of our method, from which we can observe the following findings: 
(a) The performance drops in \underline{\textit{w/o Exploration Reward}}, demonstrating that incorporating exploration rewards facilitates more effective expansions during the tree search process. 
(b) The performance drops in \underline{\textit{w/o Retrieval Reward}}, demonstrating incorporating exploration rewards enables the model to better analyze the benefits of external retrieval information during the MCTS process, thereby achieving improved comprehensive information gathering.
(c) The performance drops in \underline{\textit{w/o Progress Feedback}}, underscoring the necessity of incorporating textual feedback to guide subsequent explorations and dynamically updating the checklist.
(d) The performance drops in \underline{\textit{w/o Checklist}}, demonstrating that incorporating the explicit checklist can effectively guide the expansion and reward modeling in MCTS.
(e) The performance significantly drops in \underline{\textit{w/o HG-MCTS}}, demonstrating that the proposed HG-MCTS plays a pivotal role in enhancing the effectiveness of information seeking.

\begin{table}[t]
    \centering
    \small
\scalebox{1.03}{
    \begin{tabular}{lcccc}
        \toprule
        \textbf{Method} & \textbf{Acc.} & \textbf{R-1} & \textbf{R-2} & \textbf{R-L}  \\
        \midrule
        Ours  & 58.38 & 55.02 & 35.45 & 49.40 \\
        \midrule
        w/o Exploration Reward & 57.23 & 54.20 & 34.91 & 48.87  \\
        w/o Retrieval Reward & 56.39 & 53.68 & 34.06 & 48.15  \\
        w/o Progress Feedback & 54.57 & 52.94 & 33.73 & 47.89  \\
        w/o Checklist & 55.03 & 53.41 & 33.85 & 48.03  \\
        w/o HG-MCTS & 52.55 & 53.25 & 32.74 & 47.82  \\
        \bottomrule
    \end{tabular}}
    \caption{Evaluation results of the proposed method's variants on FanoutQA.}
    \label{tab:ablation}
\end{table}

\begin{figure}[t!]
    \centering
    \subfigure[w/o HG-MCTS]{
        \includegraphics[width=0.38\linewidth]{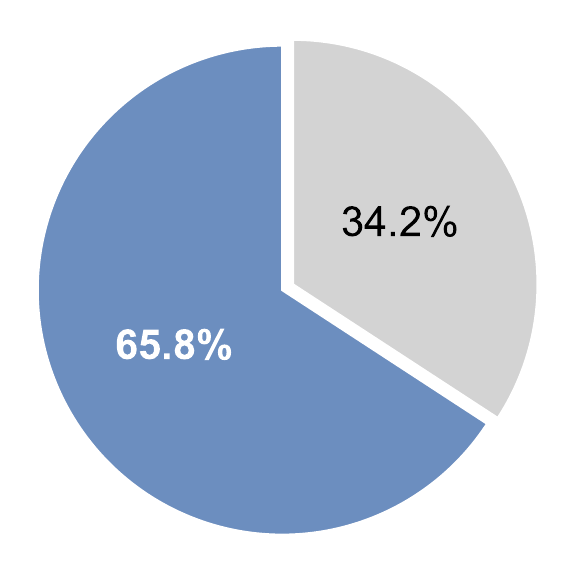}
        \label{fig:subfig1}
    }
    \hspace{0.02\textwidth} 
    \subfigure[HG-MCTS]{
        \includegraphics[width=0.38\linewidth]{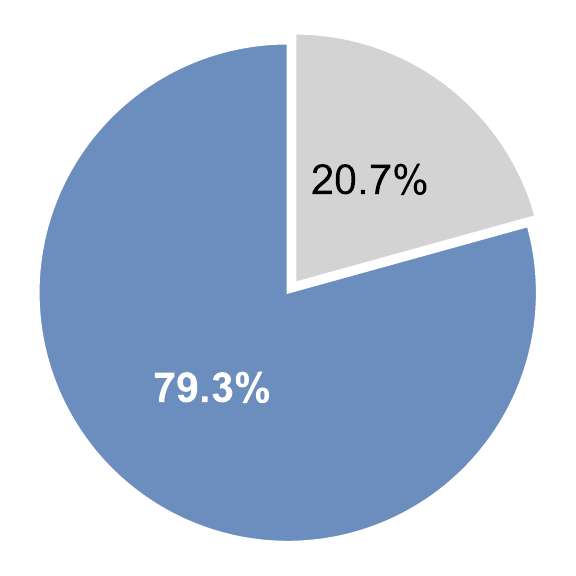}
        \label{fig:subfig2}
    }
    \caption{Information collection evaluation for different methods on Recall rate~(blue part).}
    \label{fig:recall}
\end{figure}

\subsection{Analysis on Information Collection}
To evaluate the comprehensiveness of information collected by the proposed HG-MCTS method during the reasoning process, we conducted a systematic comparison of different methods. Here, we compare the proposed method with the w/o HG-MCTS variant in Section~\ref{sec:ablation}, which also employs an information-gathering process.
Specifically, we evaluate the comprehensiveness by calculating the recall rate of the set of web pages retrieved by different methods for the ground-truth Wikipedia pages annotated as solving each intricate query. 
The ground truth serves as a benchmark, representing the authoritative and comprehensive sources necessary for addressing the query. By analyzing the coverage of the retrieved web pages relative to the ground truth, we aimed to quantify the ability of each method to gather a sufficient and relevant body of information.

As shown in Figure~\ref{fig:recall}, our method demonstrates a higher recall rate compared to the baseline in information collection evaluation, indicating that the proposed targeted MCTS enables a more comprehensive collection of information relevant to user queries. Furthermore, our method avoids the tendency to indiscriminately gather excessive information in pursuit of a higher recall rate. Such an approach, while potentially increasing coverage, introduces substantial noise into the subsequent aggregation task, thereby compromising the overall effectiveness of the information seeking.

\subsection{Scaling Law on Simulation Amount}

Our experiments systematically explore the impact of varying the number of simulation iterations on the performance of our analysis method. We adopt two LLM backbones GPT-4o-mini and DeepSeek-V3-chat on the FanoutQA dataset with various simulation numbers for analysis. 

\begin{figure}[t!]
    \centering
    \subfigure[GPT-4o-mini]{
        \includegraphics[width=0.4772\linewidth]{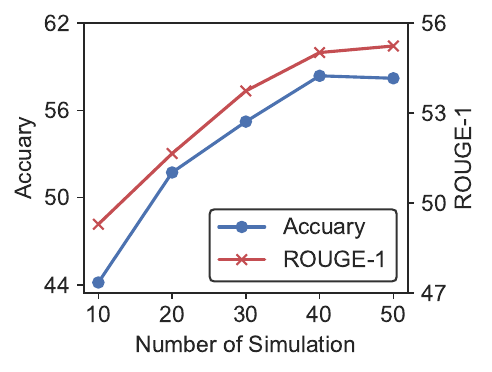}
        \label{fig:subfig1}
    }
    \subfigure[DeepSeek-V3-Chat]{
        \includegraphics[width=0.4772\linewidth]{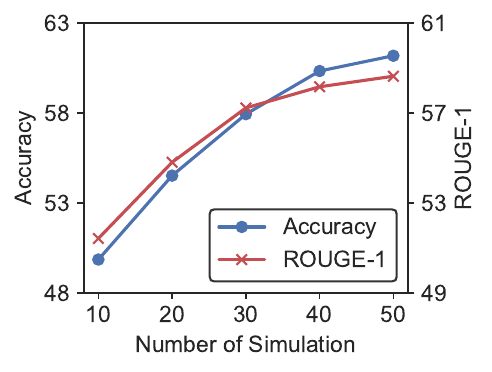}
        \label{fig:subfig2}
    }
    \caption{Evaluation results of HG-MCTS with various simulation numbers employed by different LLMs on FanoutQA.}
    \label{fig:simulation}
\end{figure}

As shown in Figure~\ref{fig:simulation}, we find that increasing the simulation count initially yields significant gains in the model’s ability to navigate the solution space effectively. With more iterations, the method benefits from a broader exploration of potential reasoning paths, leading to enhanced accuracy and improved recall in downstream tasks. In particular, the enhanced exploration reduces the likelihood of early-stage errors propagating through subsequent steps, thereby reinforcing the overall integrity of the search process.
Moreover, our results also reveal a point of diminishing returns. Beyond a certain number of simulations, additional iterations contribute only marginal improvements to the final performance. This saturation effect can be attributed to the inherent limitations of LLM's internal knowledge and the increased computational overhead, which together may lead to overly complex decision paths without proportional benefits. Thus, while extended simulations promote a more thorough examination of the search space, they also impose a trade-off between improved performance and computational efficiency.

\section{Related Work}
\label{sec:related}

\subsection{Test-Time Slow Thinking with LLMs}

The integration of the slow thinking paradigm~(\aka System 2) inspired reasoning techniques into LLMs has emerged as a pivotal research area~\cite{sutton2019bitter, wang2024q, kahneman2011thinking}, focusing on enhancing the problem-solving ability and the interoperability of LLM. A notable example is OpenAI’s o1 model\footnote{https://openai.com/o1/}, which incorporates extended internal reasoning chains during inference. 
This breakthrough has demonstrated remarkable success on programming and complex scientific benchmarks, which improve the focus of test-time reasoning that leverages extended inference processes to simulate deliberate, stepwise problem-solving akin to human-like cognitive processes without additional training~\cite{zhang2024llama, putta2024agent, luo2024improve}. Techniques such as chain-of-thought~\cite{wei2022chain} and self-consistency decoding~\cite{wangself} exemplify this paradigm, where generating intermediate reasoning steps or exploring multiple solution paths improves reliability and interoperability.
Recent advancements further extend these ideas by integrating search-based algorithms, such as beam search~\cite{kang2024mindstar} and Monte Carlo Tree Search (MCTS)~\cite{zhoulanguage, chen2024alphamath, zhang2024rest}, to systematically explore alternative reasoning trajectories. By exploring multiple outcome branches during inference, search-based methods achieve a favorable exploration-exploitation trade-off and have been widely adopted in reinforcement learning~\cite{hart1968formal, silver2017mastering} and real-world systems such as AlphaGo~\cite{silver2016mastering}. These methods are often guided by reward models, which provide feedback based on procedural or outcome-driven metrics to improve reasoning quality~\cite{snell2024scaling} iteratively. 
These collective efforts underscore a paradigm shift in LLM research, highlighting the complementary relationship between training-time strategies and scalable test-time reasoning mechanisms.

\subsection{Web Search with Complex Reasoning}

Recent advances in LLMs have shifted web search beyond simple query-response paradigms to sophisticated methods capable of multi-step reasoning for complex information access~\cite{chen2024mindsearch, reddy2024infogent}. These approaches leverage generative models to integrate and interpret diverse information sources in real time. Such capabilities are particularly critical for intricate queries that require synthesizing fragmented or context-sensitive data, where conventional search systems often fail to maintain coherence or overlook essential insights~\cite{hoveyda2024aqa, khotdecomposed}.
Initial efforts, such as WebGPT~\cite{nakano2021webgpt}, follow an iterative process of query generation, information retrieval, summarization, and synthesizing information in response to user queries.
Building on this foundation, subsequent studies adopt chain-of-thought reasoning to decompose complicated tasks into more manageable subqueries, enabling stepwise verification and refinement~\cite{yao2023react, trivedi2023interleaving}. For instance, Search-in-the-Chain~\cite{xu2024search} systematically disaggregates complex information-seeking queries by iteratively generating partial hypotheses and validating them against web-based evidence. 
Additionally, multi-agent collaboration has been explored to further enhance the search process. 
MindSearch~\cite{chen2024mindsearch} employs a planner-searcher architecture to conduct planning based on directed acyclic graphs and to carry out hierarchical information seeking. Similarly, Infogent~\cite{reddy2024infogent} introduces a multi-module collaborative agent framework that orchestrates information aggregation across modular components.
These systems excel in dynamically managing query reformulation, evidence synthesis, and inferential reasoning, seamlessly adapting to emerging information during the search process.
Moreover, some studies incorporate explicitly retrieved external information into the MCTS reasoning process, enhancing the deliberate reasoning capabilities of LLMs in multi-hop problem-solving~\cite{lee2024zero, jiang2024rag}. In contrast to these approaches, our method reformulates the task as an information collection process with the introduction of a novel checklist-based planning mechanism to holistically guide the MCTS reasoning process. Furthermore, we combines quantitative progress reward and qualitative progress feedback in reward modeling, making the MCTS process more intelligent to explore more efficient reasoning paths.

\section{Conclusion}
In this paper, we introduced HG-MCTS, a new paradigm for intricate information seeking that unites an adaptive checklist and multi-perspective reward modeling within an MCTS framework. By generating a checklist for each complex query with an explicit set of sub-goals and reinforcing the collection process, our approach systematically addresses every relevant facet of multifaceted information needs by maintaining the knowledge memory. Furthermore, the proposed multi-perspective reward modeling provides both quantitative reward and qualitative feedback for updating the tree nodes and guiding future exploration. Experimental results on multiple real-world intricate information seeking tasks show that HG-MCTS not only acquires more thorough knowledge collections but also produces more precise final responses when compared to existing baselines. 
In future work, we plan to develop fine-tuning methods for reward modeling to capture user feedback or preferences in real time, which may refine the system’s adaptability, particularly in dynamic or user-driven retrieval settings.


\bibliographystyle{ACM-Reference-Format}
\bibliography{sample-base}



\end{document}